\documentclass[CEJP,PDF]{cej}
\usepackage{layout}
\usepackage{amsmath}
\usepackage{textcomp}
\usepackage{hyperref}

\newcommand{\be}{\begin{equation}}
\newcommand{\ee}{\end{equation}}
\newcommand{\bea}{\begin{eqnarray}}
\newcommand{\eea}{\end{eqnarray}}
\newcommand{\eq}{\begin{equation}}
\newcommand{\eqx}{\end{equation}}
\newcommand{\eqn}{\begin{eqnarray}}
\newcommand{\bi}{\begin{itemize}}
\newcommand{\eqnx}{\end{eqnarray}}
\newcommand{\ei}{\end{itemize}}
\newcounter{hran}

\newcommand{\ba}{\begin{array}}
\newcommand{\ea}{\end{array}}
\newcommand{\balg}{\begin{align}}
\newcommand{\ealg}{\end{align}}

\newcommand{\lsim}
{\raise0.3ex\hbox{$\;<$\kern-0.75em\raise-1.1ex\hbox{$\sim\;$}}}
\newcommand{\gsim}
{\raise0.3ex\hbox{$\;>$\kern-0.75em\raise-1.1ex\hbox{$\sim\;$}}}

\title{Initial conditions for inflation and the energy scale of SUSY-breaking from the (nearly) gaussian sky}

\articletype{Research Article}

\author{Luis \'Alvarez-Gaum\'e\inst{1},\email{luis.alvarez-gaume@cern.ch},
            C\'esar G\'omez\inst{2}$^,$\inst{3}\email{cesar.gomez@uam.es}, 
            Raul Jimenez\inst{4}$^,$\inst{1}\email{raul.jimenez@icc.ub.edu}}

\institute{
\inst{1} Theory Group, Physics Department, CERN, CH-1211, Geneva 23, Switzerland
\inst{2} Arnold Sommerfeld Center for Theoretical Physics, Department fur Physik, Ludwig-Maximilians-Universitat Munchen Theresienstr. 37, 80333 Munchen, Germany
\inst{3} Physics Department and Instituto de Fisica Teorica UAM/CSIC, 28049 Cantoblanco, Madrid, Spain
\inst{4} ICREA \& ICC, University of Barcelona (IEEC-UB), Marti i Franques 1, Barcelona 08028, Spain}

\abstract{We show how general initial conditions for small field inflation can be obtained in multi-field models.  This is provided by non-linear angular friction terms in the inflaton that provide a phase of non-slow-roll inflation before the slow-roll inflation phase. This in turn provides a natural mechanism to star small-field slow-roll at nearly zero velocity for arbitrary initial conditions. We also show that there is a relation between the scale of SUSY breaking ($\sqrt f$) and the amount of non-gaussian fluctuations generated by the inflaton. In particular, we show that in the local non-gaussian shape there exists the relation $\sqrt f =  10^{13} {\rm GeV} \sqrt{f_{\rm NL}}$. With current observational limits from Planck, and adopting the minimum amount of non-gaussian fluctuations allowed by single-field inflation, this provides a very tight constraint for the SUSY breaking energy scale $\sqrt f = 3-7 \times 10^{13}$ GeV at 95\% confidence. Further limits, or detection, from next year's  Planck polarisation data will further tighten this constraint by a factor of two. We highlight that the key to our approach is to identify the inflaton with the scalar component of the goldstino superfield. This superfield is universal and implements the dynamics of SUSY breaking as well as superconformal breaking.}
\keywords{SUSY \*\ Inflation \*\ Cosmology}

\begin{document} 

\maketitle

\section{Introduction}

Recent constraints on inflation by the Planck satellite \cite{Planckinf, PlanckNG} have provided new insight on the properties of the inflaton. We know that the generation of non-gaussian fluctuations has been restricted significantly, with no detection by Planck and only upper limits reported (for the local case Planck reports $f_{\rm NL} < 14$--at 95\% confidence). Further, constraints on the non-detection of the tensor-to-scalar ratio ($r < 0.1$) have served to eliminate many candidates for the inflaton. In fact, the above ``non-detections" already point toward a model for the inflaton  in which perturbations were nearly Gaussian and most likely generated by a single-field slow-roll scalar with canonical kinetic energy; further, it seems likely the field is in the so-called "small-field" class with displacement of $\sim M$ (hereafter M is the Planck mass scale) to produce the required e-folds to explain flatness. An excellent review on this class of models can be found in Ref.~ \cite{Mukhanov:2013tua,Ring}. 

One very interesting question to be answered after the Planck results is how to set-up sufficiently general initial conditions for the inflaton in the "small-field" class, in other words: how can we have  inflation to start with nearly zero velocity at the time it will start slow-rolling in a flat potential? Here we present a general mechanism, inspired by SUSY, to do so. 

Unless one has a single-field slow-roll inflaton with canonical kinetic energy, nearly all other models produce measurable amounts of non-gaussianity \cite{Allenetal87,SalopekBond90,Falketal93,Ganguietal94,Verde:1999ij,WangKamionkowski2000, Komatsu:2001rj,Acquaviva:2002ud,Maldacena:2002vr,Bartolo:2004if} with values of the parameter that measures non-gaussianity $f_{\rm NL} >> O(1)$. Even the single-field slow-roll inflaton will produce values of non-gaussianity at the level of the tilt ($\sim n_s -1$) which might be detected in futuristic 21cm experiments that measure all modes in the current horizon. The nice feature of being able to measure non-gaussian fluctuations is that it provides all the correlators of the inflaton, thus one could construct, from observations, the effective lagrangian of the inflaton itself, very much in the fashion that is done in high-energy physics at accelerators like the LHC for the standard model of particles and beyond.

In the minimal-inflation\cite{minfI,minfII,minfIII} scenario the field $X$ that drives the exponential expansion of the Universe can often be represented at low energies by a Goldstino composite $GG$. Our main motivation to propose to identify the inflaton field with the order parameter of supersymmetry 
 breaking is guided by the fact that, independently of the particular microscopic mechanism driving supersymmetry breaking (in what follows we will
 restrict ourselves to $F$-breaking ) we can define a superfield $X$ whose $\theta$ 
 component at large distances becomes the ``Goldstino" (see \cite{volkovakulovrocek,seiberg2}).  In the UV the scalar component $x$ of $X$ is well
 defined as a fundamental field while in the IR, once supersymmetry is spontaneously broken, this scalar field may be expressed as a
two Goldstino state.  The explicit realisation of $x$ as a fermion bilinear depends on the low-energy details of the model.
In models of low-energy supersymmetry the realization of $x$ as $GG$ can be 
implemented by imposing a non linear constraint in the IR for the $X$ field of the type 
$X^2=0$. In our approach to inflation we use one real component of the UV $x$ field as the
inflaton. We assume the existence of a F-breaking effective superpotential for the $X$-superfield 
and we induce a potential for $x$ from gravitational corrections to the K\"ahler potential.  

In this paper we show that for generic trajectories of the minimal inflation model there is a level of generated non-gaussian fluctuations that depends directly on the scale of SUSY breaking. Therefore the further the value of non-gaussianity in the sky is constraint by current CMB experiments like Planck, the better we can constraint the SUSY energy scale and therefore make predictions for the feasibility of discovering SUSY at the LHC.

As a final observation we would like to make several remarks to highlight the similarities and differences between our approach and other attempts to identify the inflaton as well as the underlying dynamics of inflation. The key to our approach is to identify the inflaton with the scalar component of the goldstino superfield. This superfield is universal and implements the dynamics of SUSY breaking as well as superconformal breaking. In our approach SUSY breaking is unavoidably linked to inflation and the constraints we get for the SUSY breaking scale are partially dictated by requiring a SUGRA vacuum with zero cosmological constant. Finally our approach can be understood as similar to Higgs inflation with the important difference of using  what could be understood as the Higgs of the SUSY breaking. 

\section{Setup}

\begin{figure}
\begin{center}
\includegraphics[width=.65\columnwidth]{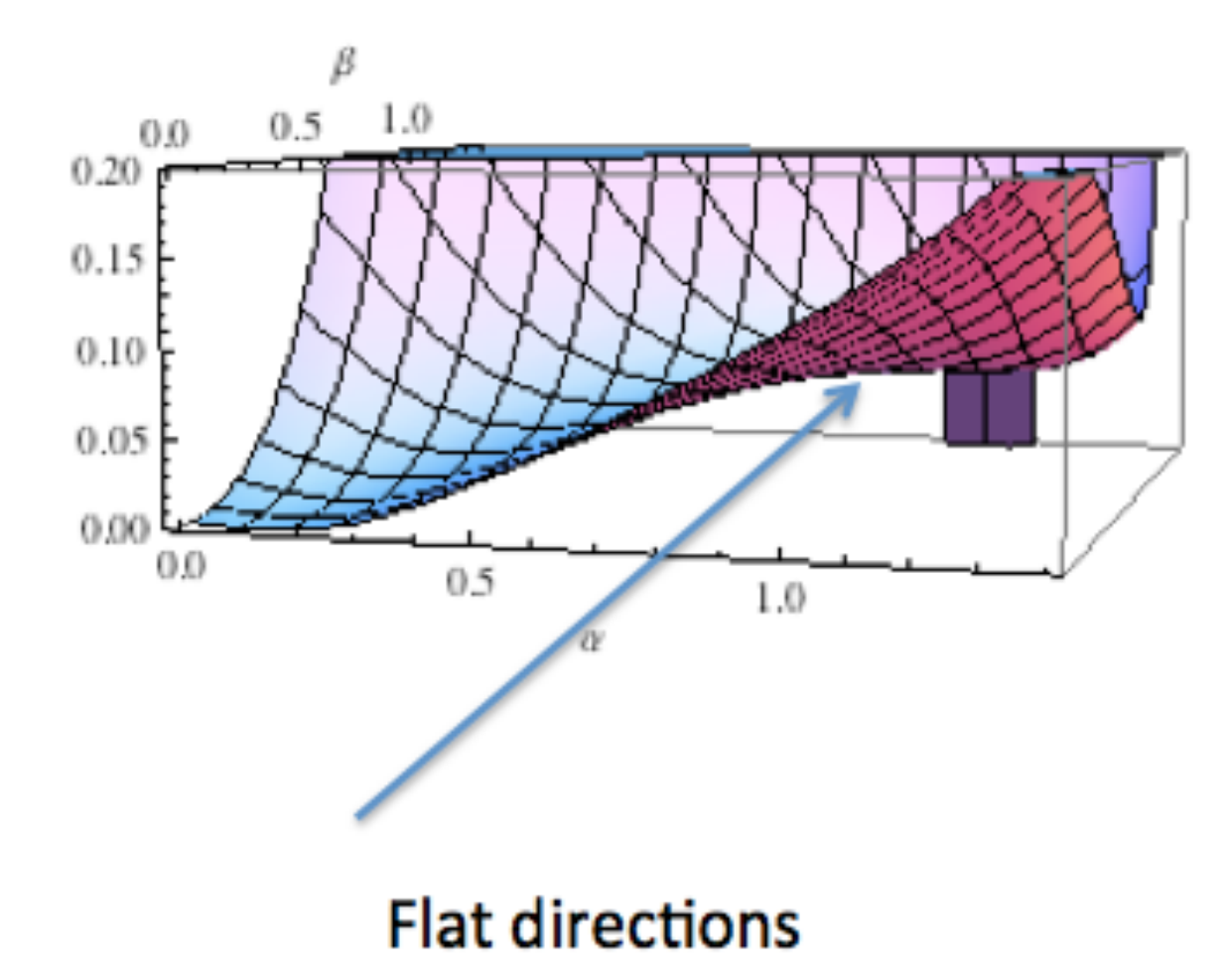}
\end{center}
\caption{The potential (shown in units of $V/f^2$) as a function of the real fields ($\alpha,\beta$) for the case with parameters ($a = 0, b = 1, c = -1.4, f_0 \sim -f$), note that this creates a very flat "bottom-valley" region, where slow-roll conditions are fulfilled.}
\label{fig:1}
\end{figure}

In order to study under what conditions the inflaton in our model will produce general initial conditions for small-field slow-roll inflation, we generate randomly $1000$ trajectories for different starting points in the inflation potential (see Fig.~\ref{fig:1}).

Let us briefly recall the form of the inflation potential and how inflationary trajectories are found.

In our minimal inflationary scenario \cite{minfI,minfII} we use only the Ferrara-Zumino (FZ)
multiplet to drive inflation.  The scalar potential in the Eisntein frame is given by:
\begin{equation}
V_{E} = e^{\frac{K}{M^2}}( -\frac{3}{M^2}W \bar W +G^{X\bar X} D_XW \, D_{\bar X}\bar W),
\end{equation}
where the K\"ahler metric and the K\"ahler covariant derivatives are given by:
\begin{equation}
G_{X\bar X}\,=\,\partial_X \bar \partial_X\, K(X,\bar X)\qquad
D\,W(X)\,=\,\partial_X\, W(X)\,+\,{\frac{1} {M^2}}\,\partial_X\,K\, W(X).
\end{equation}
In our approach we make an explicit, but reasonably generic choice for $K$ and $W$.

For us the inflaton superfield is the FZ-chiral superfield $X = z +\sqrt{2}\,\theta\psi + \theta^2 F$,
the order parameter of supersymmetry breaking. We will consider the simplest superpotential implementing
F-breaking of supersymmetry.  More elaborate superpotentials often reduce to this one once heavy fields
are integrated out.
\begin{equation}
W = f X + f_0\, M 
\end{equation}
with $f_0$ some constant to be fixed later by imposing the existence of a global minimum with 
vanishing cosmological constant and with $f$ the supersymmetry breaking scale $f= \mu_{susy}^2$.

We are interested in sub-planckian inflation, and not in the ultraviolet complete theory
that should underlie the scenario.  Hence we simply parametrize the subplanckian theory
in terms of the previous superpotential, and a general K\"ahler potential whose coefficient
will be taken of order one.  We try to use our ignorance of the ultraviolet theory to
our advantage. The K\"ahler potential $K$ we consider is:
\begin{equation}
K= X \bar X + \frac{a}{2 M} (X^2\,\bar X + c.c.) - \frac{b}{6 M^2}(X \bar X)^2
-\frac{c}{9 M^2} (X^3 \bar X + c.c.)+\ldots\,-2\,M^2\,
\log(1 + \frac{X +\bar X}{M})
\end{equation}

Which can be understood as a taylor series expansion of all terms up-to $1/M^2$ plus a term (the $\log$) that breaks R-symmetry.
In our case, the scalar fields form a complex scalar field, the partner of the goldstino field.  
Our complex field can be written as $z = M (\alpha + i \beta)/\sqrt{2}$. 
The potential is $V(z, \bar z) = f^2 V(\alpha, \beta)$, since we only include for simplicity two scales, 
the Planck scale M and the  supersymmetry breaking scale $f^{1/2}$.  In supergravity models, the gravitino
mass is up to simple numerical factors given by $m_{3/2}\sim f/M$.
It is convenient to write down dimensionless equations of motion such that time 
is counted in units of $f^{-1/2}$.  

The system of differential equations for the trayectory becomes:
\begin{eqnarray}
\alpha'' + 3 \frac{a'}{a} \alpha' + \frac{1}{2} \partial_{\alpha} \log g (\alpha'^2 - \beta'^2) + \partial_{\beta} \log g \alpha' \beta' + g^{-1} V'_{\alpha}  & = & 0 \nonumber \\
\beta'' + 3 \frac{a'}{a} \beta' + \frac{1}{2} \partial_{\beta} \log g (\beta'^2 - \alpha'^2) + \partial_{\alpha} \log g \alpha' \beta' + g^{-1} V'_{\beta} & = & 0 \nonumber \\
\frac{a'}{a} = \frac{H}{m_{3/2}} = \frac{1}{\sqrt{3}} \left ( \frac{1}{2} g  (\alpha'^2 + \beta'^2) + V(\alpha, \beta) \right )^{1/2} &  &
\label{eq:eom}
\end{eqnarray}

The coefficients $f_0, a, b, c$ will be chosen appropriately so that we obtain flat directions. 
For our purposes it is convenient to chose $a=0$, as this guarantees the existence of a global 
minimum. $f_0$ will be adjusted such that  the global minimum is at a vanishing value
of the potential. So the model only has $b,c$ 
as free parameters, which we try to keep of order one to avoid fine-tunning in the potential. 

From the collection of potentials considered, not all will show flat directions 
where it is possible to inflate during enough e-foldings ($> 55$) for any choice of the 
microscopic parameters $f_0, a, b, c$. We will restrict to cases where the potential has a global 
minimum with vanishing cosmological constant and thus we fix 
the value of the minimum at $0$ tuning the value of $f_0$.

\begin{figure}
\begin{center}
\includegraphics[width=.45\columnwidth]{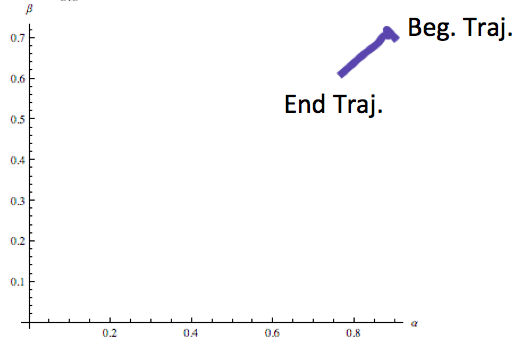}
\includegraphics[width=.45\columnwidth]{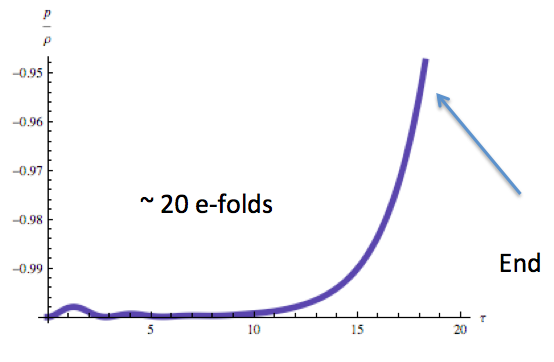}
\end{center}
\caption{Left panel: a trajectory in the plane $\alpha, \beta$ for zero velocity initial conditions. Right panel: the equation of state as a function of e-folds. Note that trajectory stays always in the slow-roll regime for nearly 20 e-folds.}
\label{fig:2}
\end{figure}

\begin{figure}
\begin{center}
\includegraphics[width=\columnwidth]{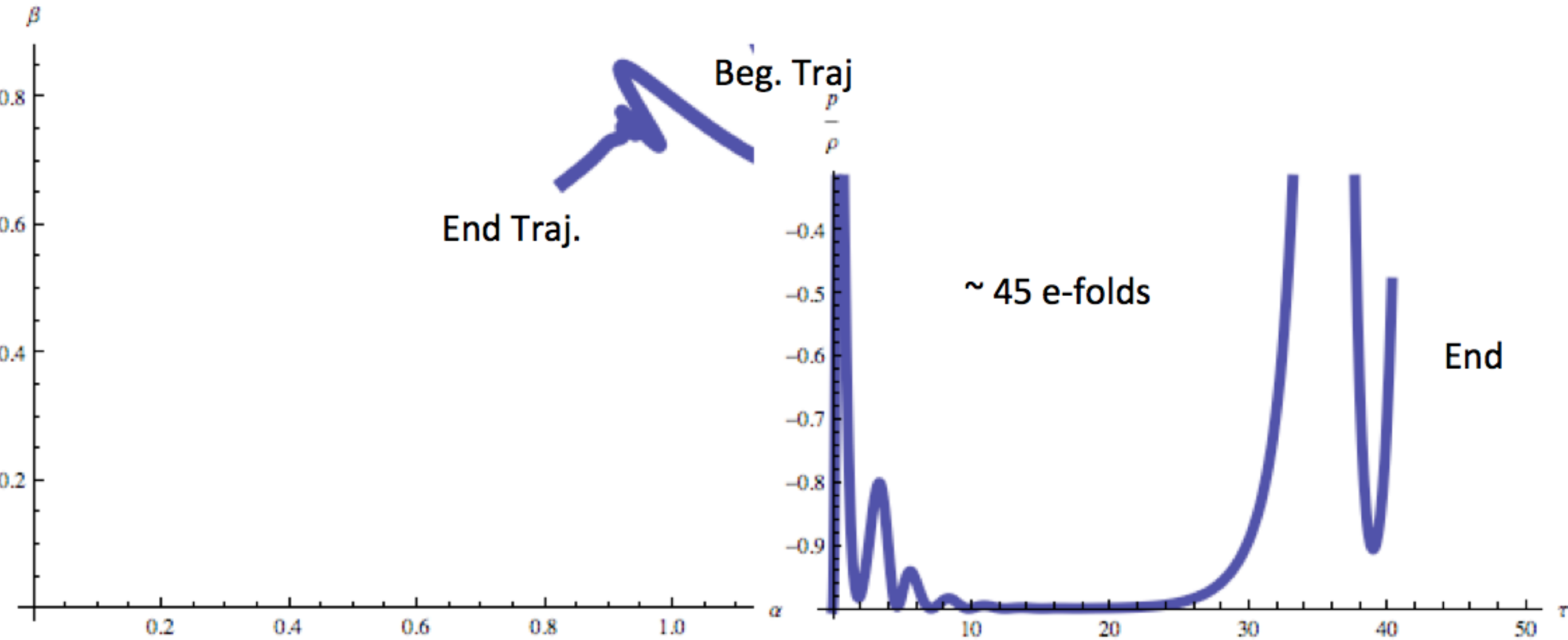}
\end{center}
\caption{An example of a trajectory for the potential shown in Fig.~\ref{fig:1}. The left panel shows the trajectory in the plane $\alpha, \beta$. Note that we start the trajectory at an arbitrary position with random velocities in the steep part of the potential. Because of the non-linear friction in our equations of motion, every time the inflaton turns, there will be significant friction and therefore will be slowed down. This happens until it reaches the valley where it slow-rolls as a single field inflaton. The right panel shows the equation os state ($p/\rho$) as a function of number of e-folds ($\tau$). Recall that $p/\rho \sim (\epsilon-1)$, where $\epsilon$ is the first slow-roll parameter, so that traditional slow-roll happens when $p/\rho = -1$. The trajectory generates non-gaussianity both at the largest scales (at the current horizon scale) and the smallest (below the dwarf galaxies). It is in these scales that non-gaussianity could be generated depending on the SUSY scale.}
\label{fig:3}
\end{figure}

We can now compute trajectories and attractors in more detail using the equations above. An example of the potential is shown in Fig.~\ref{fig:1} for values of the parameters $a=0, b=1, c=-1.5$. Note that there is a flat direction where slow-roll takes place. We elaborate on this in the next section.

\section{General initial conditions for small-field inflation}

\begin{figure*}
\includegraphics[width=3in]{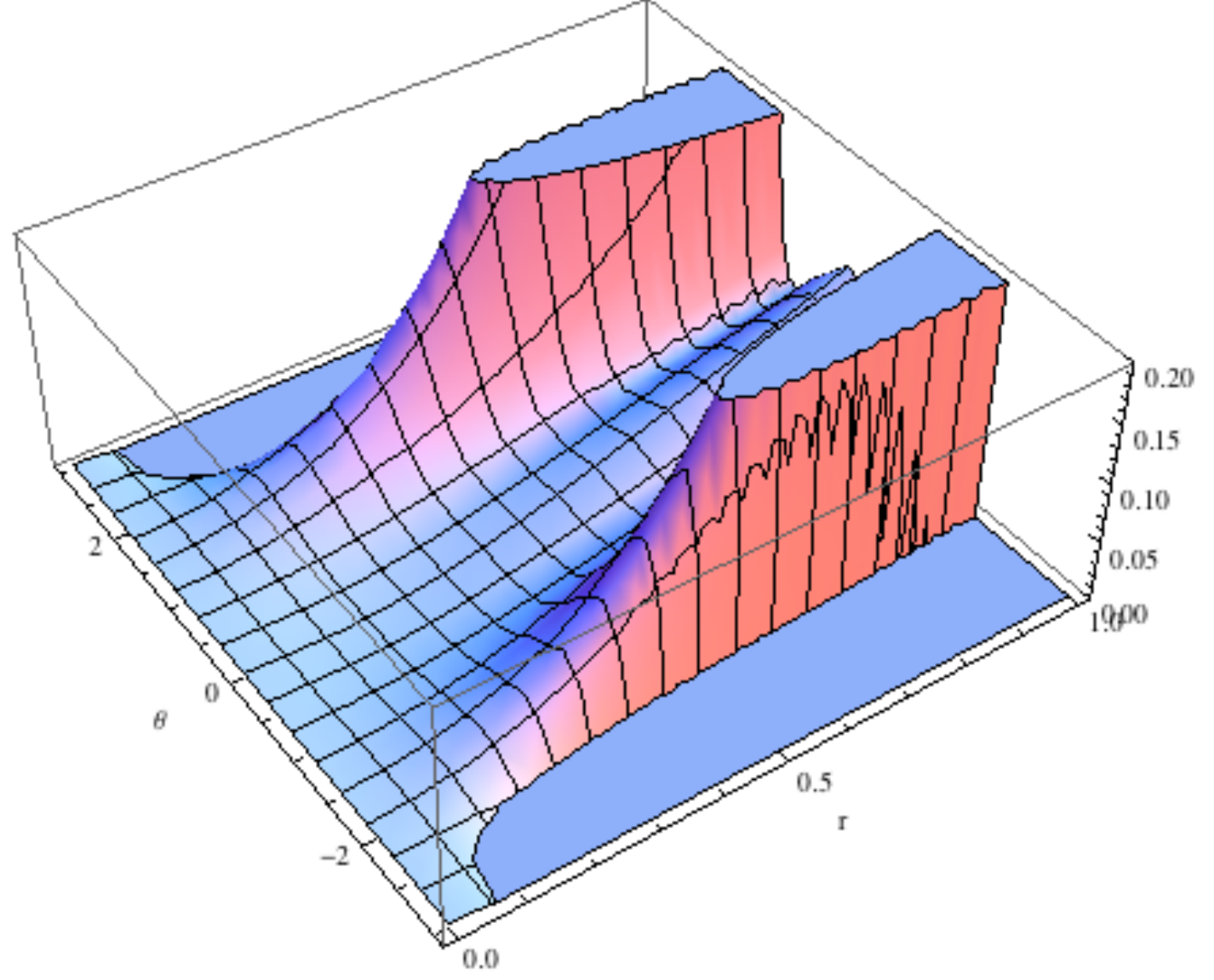}
\includegraphics[width=3in]{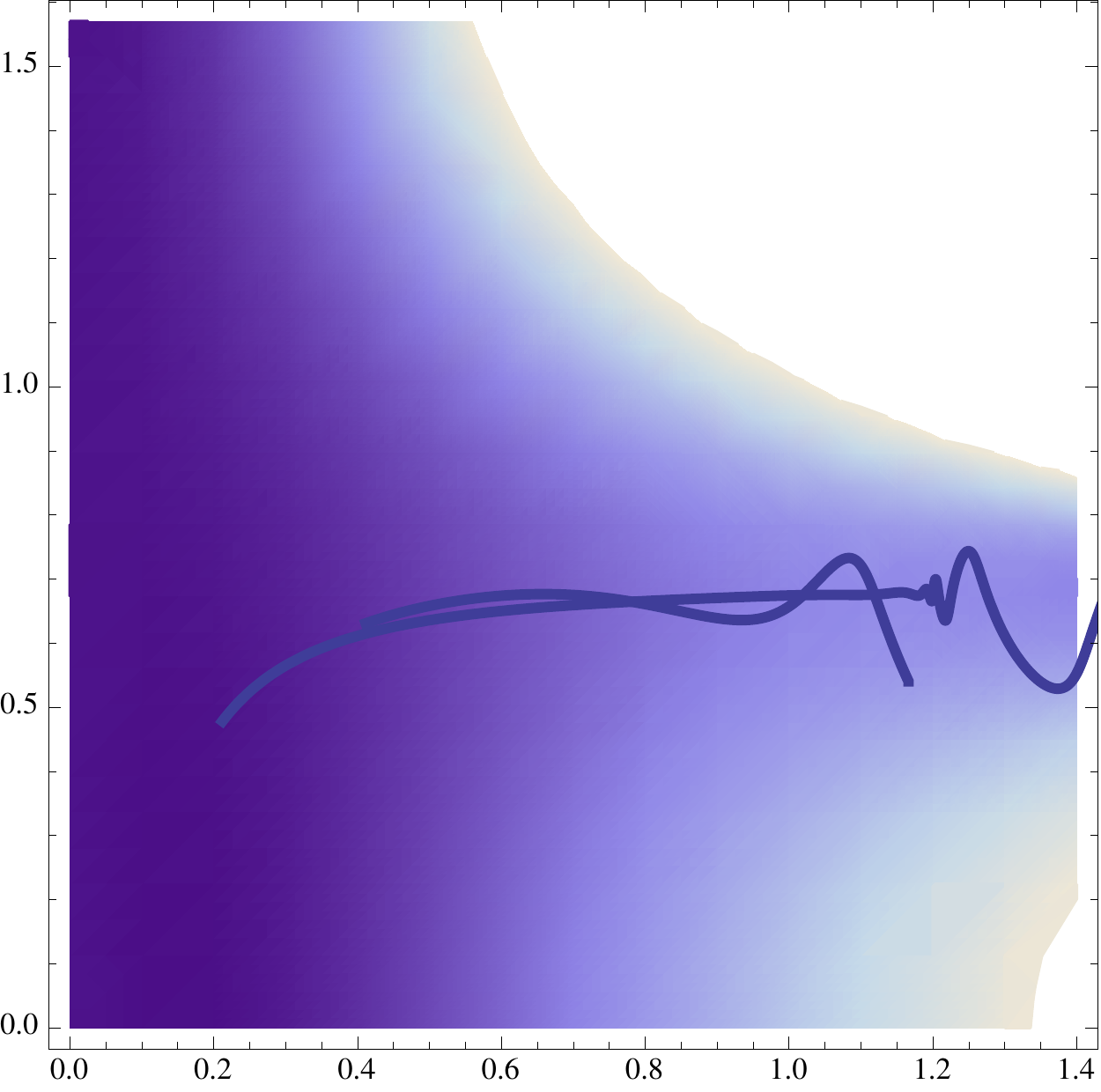}
\caption{Left panel: polar plot of the potential used to compute inflationary trajectories. Right panel: contour plot version of the left panel ($\theta$ (y-axis) range from $0$ to $\pi/2$) with some inflationary trajectories over-plotted.}
\label{fig:cont}
\end{figure*}

There are a number of important properties of the system of equations~\ref{eq:eom} and its
solutions that are shared by large classes of supersymmetric theories.  In our approach, 
the inflaton is always the scalar component of the goldstino superfield, and the
K\"ahler potential and the superpotential completely determine its dynamics. For a 
multifield inflationary theory, with a non-canonical kinetic term,
the dynamical equations of motion take the form:
\begin{eqnarray}\nonumber
H^2 & = & {8\,\pi\,G\over 3}\left({1\over 2} G_{ij} \dot{X}^i\,\dot{X}^j\,+\,V(X)\right) \\ 
{D\dot{X}^i\over dt} & + & 3\,H\,\dot{X}^i\,+\,G^{ij}\,\partial_j\,V\,=\,0, 
\end{eqnarray}
where $D/dt$ is the covariant derivative with respect to the metric $G$ in field space.
We can define the energy functional for a given trajectory as:
\begin{equation}
E[X]\,=\,{1\over 2} G_{ij} \dot{X}^i\,\dot{X}^j\,+\,V(X).
\end{equation}
It is easy to see that in an expanding universe it is a monotonically decreasing function of
time:
\begin{equation}
{d\,E[X]\over d t}\,=\,- 3\, H\, G_{ij} \dot{X}^i\,\dot{X}^j
\end{equation}
When we use the scalar component of the goldstino superfield, the  trajectories are
always plane curves in the plane $(\alpha,\beta)$ whose metric is always of Gaussian
form:
\begin{equation}
ds^2 \,=\,2 G(\alpha,\beta)\,( d\alpha^2\,+\,d\beta^2),
\end{equation}
and the first two equations in~\ref{eq:eom} can be succinctly written as:
$$
\ddot{z}\,+\,\partial_z\log G\,\dot{z}^2\,+\,3\,H\,\dot{z}+ G^{-1}\,\partial_{\bar{z}}\,V\,=\,0.
$$
The slow roll equations are simply:
$$
3\,H\,\dot{z}+ G^{-1}\,\partial_{\bar{z}}\,V\,=\,0.
$$
It is clarifying to analyse these equations (the first and second order sets) in terms
of polar coordinates.  In the models we consider we start inflation close but below
the Planck scale.  In polar coordinates, the change of the energy of the system is given
by:
$$
{d\,E[X]\over d t}\,=\,- 3\, H\, G(\rho,\theta)\,(\dot\rho^2+\rho^2\,\dot{\theta}^2),\qquad
z\,=\,\rho\,e^{i\,\theta}
$$
The initial value for $\rho$ will be close to one in Planck units, hence if we choose
some generic initial conditions with an arbitrary direction for the speed of the field,
it is clear that for high values of $\dot\theta$ the slashing of the angular component
will rapidly damp the energy and the field will join any of the trajectories determined
by the extrema of the potential with respect to the angular variable $\partial_{\theta}\,V\,=\,0$.
From the angular part of the slow roll equations one sees easily that those constant
$\theta=\theta_0$ satisfying the extremum condition are exact solutions to the slow roll
equations.  Each of those values is a potential attractor.  General trajectories will
join one of this attractors and then the $\rho$ will roll to the origin as in 
single field inflation theories.  The number of e-folding generated will depend, of course, 
on the initial conditions and the parameters of the model, but it is important to remark
that in most models in this approach it is typical to obtain a number of order ten
e-foldings.  The number of effective attractor trajectories depends on the potential
and the metric.  
\vskip.5cm
If we analyse the full set of second order equations, hence without the slow roll conditions,
the conclusions are rather similar.  The attractor-like trajectories, i.e. exact solutions
with $\theta=\theta_0$ constant are also characterised by the extremal point of the potential
with respect to the angle.  Some examples can be found in the figures below.  For some
values of the parameters of our model, we plot the potential in polar coordinates.
It is easy to see attractor trajectories in the three dimensional plot in Fig.~\ref{fig:1},
and also the corresponding valleys of attraction in Fig.~\ref{fig:cont}, which portrays the same
potential but in a contour plot.  
\vskip.5cm
In the next section we explore some explicit examples and trajectories
with reasonable number of efoldings.  The general remarks just presented of course
apply to the cases studied below.

\section{Examples} 

As explained before, we always need to set $a=0$, so we concentrate on the values of $b$ and $c$. From a Monte-Carlo simulation that samples more that $1000$ values for $b$ and $c$ we have found that the most favourable values to produce enough e-folds is when $b \sim -c$ , so from now on we focus on this case. This is the case already depicted in Fig.~\ref{fig:1}. Note the main features of the potential: a very steep part at values of the field $\sim M$, a flat part at $< M$ and finally a global minimum. Let us start with a trajectory where the field starts near the flat part. This is shown in Fig.~\ref{fig:2}. the left panels shows the trajectory in the plane $\alpha, \beta$ while the right panel shows the value of the equation of state $p/\rho$, recall that slow-roll implies $p/\rho \sim -1 \sim (\epsilon-1)$ where $\epsilon$ is the first slow-roll parameter, as a function of the number of e-folds ($\tau$). First, the trajectory is very flat (note the small values of $p/\rho$ on the y-axis) but last only for about $20$ e-fold, enough to explain the observed universe but not its flatness. 

So we now explore the case when the field starts from the steeper part at positions of the field $\sim M$ and with arbitrary position in velocity and direction. An example is shown in Fig.~\ref{fig:3}. First, note that despite the initial steepness of the potential and arbitrary velocity, the field is slowed-down by the non-linear friction terms at the turns. This period os ``slashing" ends into the field reaching nearly zero velocity at the beginning of the flat part of the potential. From the right panel we observe that this early phase provides a few e-folds before entering the slow-roll phase that last for about $25$ e-folds. The field then exits slow-roll and enters a final phase of no slow-roll adding and extra 10 e-folds. In total we obtain the required 45 e-folds to explain flatness and the required slow roll phase to explain the observed slope of the primordial power spectrum $n_s = 0.96 \pm 0.007$. 

This is typical of what we found in the Monte-Carlo simulation for arbitrary trajectories starting at positions of the field $\geq M$. Note that because we do not have control on how the potential behaves beyond values of the field of $M$, it is important that even for very steep values the non-linear friction is efficient at slowing down the inflaton and providing initial conditions for slow-roll. 

Thus general trajectories in our model look very much like the one in Fig.~\ref{fig:3}. Note that obtaining order $50$ e-folds is not difficult with values of $b, c$ order one (as in Fig.~\ref{fig:1}). More e-folds can be obtained if the values of $b,c$ are tune in one part in 100, although this does not seem to be necessary. We note that further fine-tuning of the parameters will not lead to any improvement of the model.

\begin{figure}
\includegraphics[width=3in]{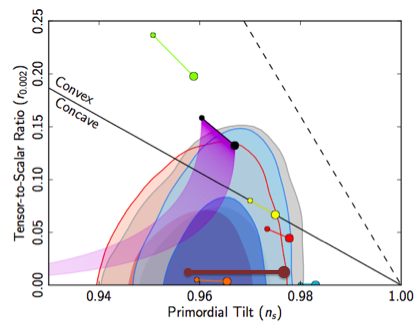}
\caption{Dark-red thick line shows the prediction for our model in the plane $r-n_s$ (plot adapted from Planck team release).}
\label{fig:r}
\end{figure}

We show the general prediction of our model for the ratio of tensor-to-scalar perturbations in Fig.~\ref{fig:r}. As expected \cite{Verde:2005ff}, because of the small displacement of the field ($\Delta \alpha, \beta \sim 0.2-0.3 M$), r is small $\sim 0.001$.

\section{Non-gaussianities}

We now answer the following question:  will non-gaussian fluctuations be generated by our model? We first note that the kinetic term is always non-canonical, but weakly so for $\alpha, \beta < 1$ as can be seen from the choice of the Khaler potential. So although our inflaton is a ``pion" we will not generate any non-gaussianity from this source. The only place where one could generate measurable non-gaussianity is from those situations in which the inflaton turns. 

We can estimate the value of the non-gaussian fluctuations following \cite{Chen:2009we}. The overall level of non-gaussianity is given by their Eq.~ 17, which reads
\begin{equation}
f_{\rm NL}^{int} = \alpha(\nu) \frac{1}{P_{\xi}^{1/2}} \left ( \frac{-V'''}{H} \right ) \left ( \frac{\dot \theta}{H} \right )^3
\label{eq:ng}
\end{equation}
where $V'''$ is the third derivative of the potential at the turn, $\dot \theta$ is the angular velocity of the inflaton as it turns, $P_{\xi}$ is the power spectrum of the fluctuations and  $\alpha(\nu)$ is a numerical factor (which we compute using \cite{Chen:2009zp}).

In order to estimate the different terms in the above equation in our models we proceed as follows: we generate $1000$ random trajectories for different values of our potential, but limited to the case where $c = -g \times b$, where $g$ is a number between $1-2$, as we know by previous experience that this is the case when the potential can harbour trajectories with $O(40-50)$ efolds; we also limit $b, c$ to have the freedom to vary only in the first decimal place as to not produce fine tunning of the potential. Finally, the trajectory are all started with random values for both position and velocity at different values of the two real fields $\alpha, \beta$. A typical trajectory is shown in Fig.~\ref{fig:3}. Note that the interesting part of the trajectory where non-gaussianities can be generated are at very large scales, comparable to the horizon scale today and at scales smaller than dwarf galaxies, i.e. very small scale perturbations. This behaviour is typical of our model for most trajectories.

\begin{figure}
\begin{center}
\includegraphics[width=\columnwidth]{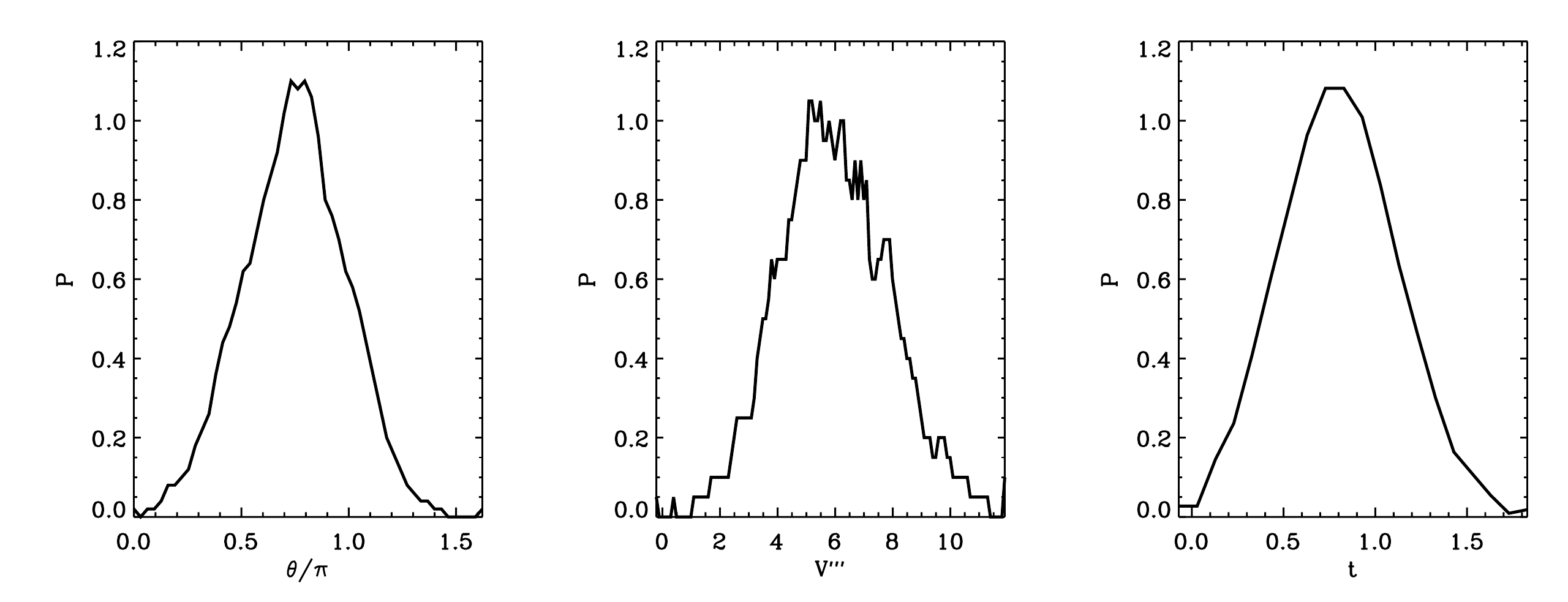}
\end{center}
\caption{The probability distribution for the values of $\theta$ (angle of the turn), $V'''$ (third derivative of the potential) and $t$ (time it takes to turn) as derived from $1000$ random trajectories generated for different initial conditions and values of the potential that generate at least 45 e-folds. These values are used to compute the amount of non-gaussianity generated in the trajectories. }
\label{fig:4}
\end{figure}

From this set of trajectories we compute the distribution of $\dot \theta$ and $V'''$. This is shown in Fig.~\ref{fig:4}. We can now evaluate Eq.~\ref{eq:ng} noting that $\alpha(\nu) \sim 10$, $\frac{\dot \theta}{H} \sim 10$, $P_{\xi}^{1/2} = 6 \times 10^{-9}$ and $\frac{-V'''}{H} \sim 5 (f/M^2)$, thus the relation between the SUSY breaking scale and the level of non-gaussianity reads
\begin{equation}
\sqrt f = \frac{\sqrt{f_{\rm NL}}}{10^5} M 
\end{equation}
Using current observational limits from Planck\cite{PlanckNG} ($f_{\rm NL} < 14$), and adopting the minimum amount of non-gaussian fluctuations allowed by single-field inflation\cite{Verde:2009hy}, provides a very tight constraint for the SUSY breaking energy scale $\sqrt f = 3-7 \times 10^{13}$ GeV at 95\% confidence.

In passing we note that the turns will generate isocurvature fluctuations at a level similar to the one needed to explain the observed power asymmetry at large scales \cite{Dai:2013kfa,Ade:2013nlj}; we will elaborate on this subject in a future publication.

\section{Conclusions}

We have presented in this article some more quantitative phenomenological findings for our proposal to identify the inflaton with the order parameter of SUSY breaking. We are motivated by finding a physical candidate for the inflaton, which seems to be the paradigm supported by current cosmological observations \cite{Planckinf} to explain the origin, size, flatness and perturbations of the Universe. The model is successful at answering questions about the fundamental physics behind inflation. In particular:

\begin{enumerate}

\item Why does the inflaton start the slow-roll phase with nearly zero velocity? Because the non-linear friction term (loss of angular momentum) provided by the fact that we have broken R-symmetry; this produces a "slashing" phase.

\item Why does the universe inflate $\sim 50$ e-folds? Because the inflaton rolls for about one Planck mass.

\item Why is the value of the CMB fluctuations the observed one? Due to the fact that in this model the fluctuations are proportional to the SUSY breaking scale, so thus the value of the fluctuations on the sky are linked to the energy chosen by nature to break SUSY.

\item Why does inflation end? At low energies the inflaton "integrates itself out" and manifests as a fermi gas of goldstinos, thus not behaving anymore as a scalar field. 

\end{enumerate} 

Our model makes a very precise prediction: that the scale of SUSY breaking has to be at $\sim 10^{13}$ GeV. This can be tested in the next LHC run starting in 2015. This relatively high energy scale implies that the possibility of observing SUSY at the LHC is slim but not completely excluded. The details will depend on the explicit parameters chosen to make contact with the low energy world ($\sim 10$ TeV).

%
%

\begin{thebibliography}{99}

\bibitem[Planck Collaboration et al.(2013)]{Planckinf} Planck 
Collaboration, Ade, P.~A.~R., Aghanim, N., et al.\ 2013, arXiv:1303.5084 

\bibitem[Planck Collaboration et al.(2013)]{PlanckNG} Planck 
Collaboration, Ade, P.~A.~R., Aghanim, N., et al.\ 2013, arXiv:1303.5082 

\bibitem{Mukhanov:2013tua}
  V.~Mukhanov,
  arXiv:1303.3925 [astro-ph.CO].

\bibitem[Martin et al.(2013)]{Ring} Martin, J., Ringeval, 
C., \& Vennin, V.\ 2013, arXiv:1303.3787 

\bibitem{Allenetal87} Allen, T.~J., Grinstein, 
B., \& Wise, M.~B.\ 1987, Physics Letters B, 197, 66 
\bibitem{SalopekBond90} Salopek, D.~S., \& Bond, J.~R.\ 1990, PRD, 42, 3936 
\bibitem{Falketal93} Falk, T., Rangarajan, R., 
\& Srednicki, M.\ 1993, ApJL, 403, L1 
\bibitem{Ganguietal94} Gangui, A., Lucchin, F., 
Matarrese, S., \& Mollerach, S.\ 1994,ApJ, 430, 447 

\bibitem{Verde:1999ij}
  L.~Verde, L.~-M.~Wang, A.~Heavens and M.~Kamionkowski,
  Mon.\ Not.\ Roy.\ Astron.\ Soc.\  {\bf 313} (2000) L141

\bibitem{WangKamionkowski2000} Wang, L., \& Kamionkowski, M.\ 2000, PRD, 61, 063504 

\bibitem{Komatsu:2001rj}
  E.~Komatsu and D.~N.~Spergel,
  Phys.\ Rev.\ D {\bf 63} (2001) 063002
  [astro-ph/0005036].

\bibitem{Acquaviva:2002ud}
  V.~Acquaviva, N.~Bartolo, S.~Matarrese and A.~Riotto,
  Nucl.\ Phys.\ B {\bf 667} (2003) 119
  [astro-ph/0209156].

\bibitem{Maldacena:2002vr}
  J.~M.~Maldacena,
  JHEP {\bf 0305} (2003) 013
  [astro-ph/0210603].
  
\bibitem{Bartolo:2004if}
  N.~Bartolo, E.~Komatsu, S.~Matarrese and A.~Riotto,
  Phys.\ Rept.\  {\bf 402} (2004) 103
  [astro-ph/0406398].



  
  \bibitem{minfI}
  L.~\'Alvarez-Gaum\'e, C.~G\'omez, R.~Jimenez,
  Phys.\ Lett.\  {\bf B690}, 68-72 (2010).
  [arXiv:1001.0010] [hep-th]]

\bibitem{minfII}
  L.~Alvarez-Gaume, C.~Gomez, R.~Jimenez,
  JCAP {\bf 1103 } (2011)  027.
  [arXiv:1101.4948 [hep-th]].
    
\bibitem{minfIII}
  L.~Alvarez-Gaume, C.~Gomez and R.~Jimenez,
  JCAP {\bf 1203} (2012) 017
  [arXiv:1110.3984 [astro-ph.CO]].  
  
  \bibitem{volkovakulovrocek}
D.~V.~Volkov and V.~P.~Akulov,
Phys.\ Lett.\  B {\bf 46}, 109 (1973).

\bibitem{seiberg2}
 Z.~Komargodski and N.~Seiberg,
  JHEP {\bf 0909}, 066 (2009)
  [arXiv:0907.2441 [hep-th]].

\bibitem{Verde:2005ff}
  L.~Verde, H.~Peiris and R.~Jimenez,
  JCAP {\bf 0601} (2006) 019
  [astro-ph/0506036].

\bibitem{Chen:2009we}
  X.~Chen and Y.~Wang,
  Phys.\ Rev.\ D {\bf 81} (2010) 063511
  [arXiv:0909.0496 [astro-ph.CO]].
  
\bibitem{Chen:2009zp}
  X.~Chen and Y.~Wang,
  JCAP {\bf 1004} (2010) 027
  [arXiv:0911.3380 [hep-th]].

\bibitem{Verde:2009hy}
  L.~Verde and S.~Matarrese,
  Astrophys.\ J.\  {\bf 706} (2009) L91
  [arXiv:0909.3224 [astro-ph.CO]].

\bibitem{Dai:2013kfa} 
  L.~Dai, D.~Jeong, M.~Kamionkowski and J.~Chluba,
  arXiv:1303.6949 [astro-ph.CO].

\bibitem{Ade:2013nlj} 
  P.~A.~R.~Ade {\it et al.}  [Planck Collaboration],
  arXiv:1303.5083 [astro-ph.CO].

\end{thebibliography}
\end{document}